# Discovery of an "Eclipse" in the WC9d-Type Wolf-Rayet Star, WR 53

**Rod Stubbings**
*Tetoora Road Observatory, 2643 Warragul-Korumburra Road, Tetoora Road 3821, Victoria, Australia; stubbo@sympac.com.au*

ABSTRACT.

The WC9d-type Wolf-Rayet star WR 53 was observed visually entering into an "eclipse" with a depth of 1.2 magnitude. Subsequent visual and CCD data showed a steady linear rise over 10 days to recover and return to its normal brightness level. This is the first-ever recorded "eclipse" of this star which has previously shown no photometric variability.

INTRODUCTION.

Wolf-Rayet stars (WR-stars) are massive, luminous stars with unusual spectra showing prominent broad emission lines of helium, nitrogen or carbon. Some Wolf–Rayet stars of the carbon sequence ("WC"), especially those belonging to the latest types, are noticeable due to their production of dust. They have lost or burnt almost all of their hydrogen and are now fusing helium in their cores, or heavier elements for a very brief period at the end of their lives.

Infrared (IR) excess due to the existence of dust in the winds of WC late-type stars was identified by Allen et al (1972), Gehrz & Hackwell (1974) and Cohen, Barlow & Kuhi (1975). The dust shells are being restored as the dust is swept away by the stellar wind. With new dust being formed continuously in some WC stars, this suggested their designation as persistent dust makers Williams, van der Hucht (2000). The classification introduced by van der Hucht (2001) noted persistent dust makers as WCd, with "d" indicating a persistent dust maker.

A totally different manifestation of dust formation by WR stars is eclipsing of the visual light by line-of-sight dust clumps that are not caused by the companion, or variability of the star itself. This model was developed by Veen et al. (1998) who suggests occasional "eclipses" shown by other dusty WC stars (like WR 103, WR 113 and WR 121) are obscurations by clouds in the line of sight. Another significant observation was the colour changes during the "eclipses" of WR 103 and WR 121. It shows the star becomes redder, caused by short term condensing dust clouds. Veen et al. (1998)

The presence of dust shells around WC late- type stars was also investigated by Williams, van der Hucht & Thé (1987) using infrared photometry, who found WR 53 to be a dust maker from its IR emission. The spectrum of WR 53 has since been reclassified from WC8d to WC9d by Crowther et al. (1998).

General information on WR stars can be found in the catalogue by Crowther (2007)

The discovery and a well-defined "eclipse" of WR 53 is presented in this paper.

HISTORY

The discovery of a probable Hi interstellar bubble associated with the Wolf-Rayet star WR 53 was found by Martin et al. (2007)

Photometric V- and I-band variability in WR 53 was not detected in a 20-day run by Fahed, Moffat & Bonanos (2009).

Photometry from ASAS3 V-band data during the years 2001-2010 showed no eclipse or fading events. (Pojmański 1997; Williams 2014).

All the surveys (ASAS-3, HIPPARCOS (Perryman *et al.* 1997), and APASS) have recorded WR 53 having V ~ 10.53 over the past decades.

OBSERVATIONS

Visual observations of WR 53 first started in 2005 by Albert Jones who noted some slight variations. The author (RS) monitored WR 53 from 2011, also showing variations. These variations have yet to be proved as past photometric data on WR 53 have shown the star to be constant.

On July 15, 2015, a visual observation on WR 53 showed it to be fainter than normal at magnitude 10.7. The previous estimate on July 7 was recorded at 10.5. The next observation was on July 19, and the star had notably faded further to magnitude 11.0. The following night a further drop in brightness to 11.5 was observed and this was clearly an "eclipse" event in progress. A request was sent privately that night to Peter Nelson to take an image of WR 53, which showed the star at V= 11.66. An alert request was then sent to observers for follow- up CCD time series observations, which were commenced the next night, along with visual observations during this event until the star returned to its normal quiescence level of V= 10.53.

The "eclipse" was first noticed on July 15, and from this point showed a fading trend of 5 days to minimum, with a depth of 1.2 magnitude. A steady linear rise of 10 days followed. The light curve is presented in (Fig. 1).

A spectrum was obtained after the "eclipse" event at the El Leoncito Astronomical Complex CASLEO (Argentina) which was similar to ones published in 2001 and 2012. (Roberto Gamen, private communication, 2015).

Data from the AAVSO Photometric All-Sky Survey (APASS; Henden *et al.* 2014) taken a month earlier on JD 2457163.2416 (May 20, 2015) showed a (B-V) of 0.45. Both V and B bands were observed during the egress of this "eclipse". From the AAVSO database

(Kafka 2015) on JD 2457225.4733 (July 21) the (B-V) was 0.52 just after minimum. On JD 2457236.4693 (August 1), at the end of the "eclipse" the (B-V) was 0.45. This shows a slight reddening during minimum. The light curve is presented in (Fig. 2).

SUMMERY

WR 53 has been constantly monitored visually by the author since 2011 with no sign of deep fading episodes. ASAS3 V-band data during the years 2001-2010 showed no eclipse or fading events by Williams (2014). There was a dip in 2003 around 10.8 but the data showed no faint observations either side of this. Another dip from the data in 2008, again around 10.8, showed no observations a few days before and up to a week later. Given the discovered "eclipse" activity lasted for over 15 days and at a depth of 1.2 magnitude, this dip also seems unlikely. Of course, there are the usual seasonal gaps in the observing where shallow "eclipses" could have been missed.

WR 53 is one of only a few of the late-type WC stars that has shown a deep "eclipse", WR 104 (2.7 mag.) and WR 103 (1.3 mag.) are others. Typically, their durations can last up to several weeks as shown by Veen et al. (1998). Apart from WR 104 and WR 106, which show frequent "eclipses" of varying depths, other WC8 and WC9 stars have only shown shallow "eclipses" after years of monitoring Williams (2014). What becomes apparent is how rare these "eclipse" events are in WC late-type stars.

An intensive monitoring program of WR 53 is needed to find more "eclipse" events in the hope that spectra and photometric multi-colour monitoring can be taken when this occurs.

The present discovery of this "eclipse" event in WR 53 has established its variability and has now been included as a variable star in the International Variable Star Index (VSX) (Watson, et al. 2015)

ACKNOWLEDGEMENTS

It is a pleasure to thank P. M. Williams for his comments. This research has made use of the International Variable Star Index (VSX) database, operated at AAVSO, Cambridge, Massachusetts, USA, and through the use of the AAVSO Photometric All-Sky Survey (APASS), funded by the Robert Martin Ayers Sciences Fund. I would like to thank Peter Nelson for providing a V- band image when faint, Josch Hambsch and Rolf Carstens for follow-up CCD observations. I would also like to thank the anonymous referee whose comments and suggestions were very helpful.

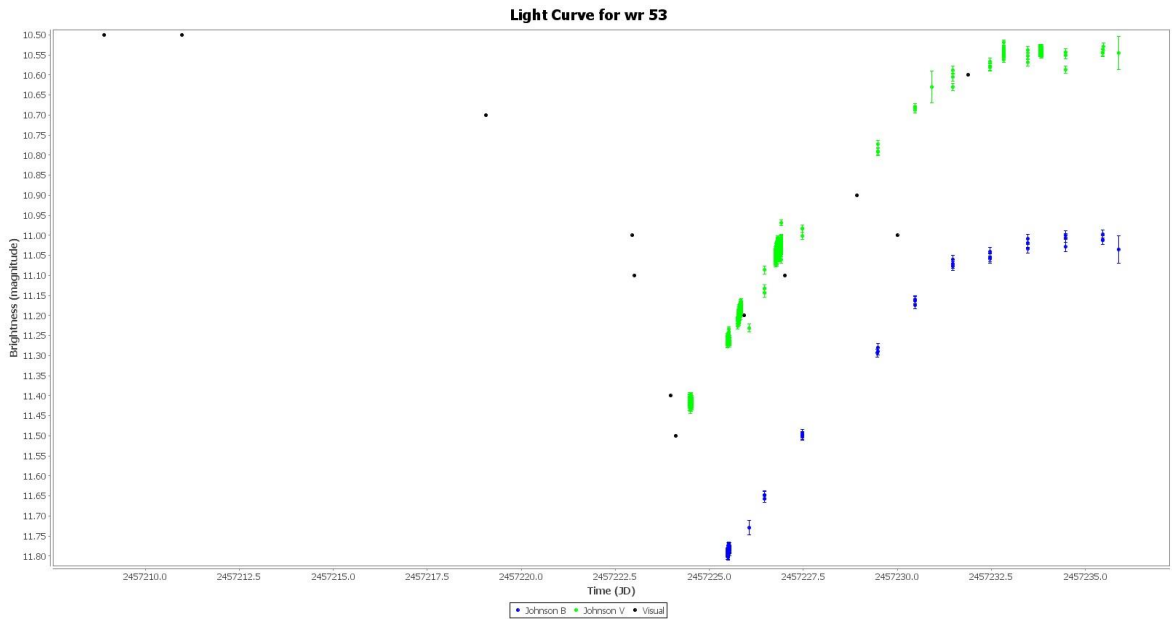

Figure 1. Light curve of WR 53. AAVSO data from JD 2457208 through JD 2457236 (July 4, 2015 - August 1, 2015) showing the first ever "eclipse" of the Wolf-Rayet star WR 53. Black dots are visual data, green are Johnson V and blue are Johnson B.

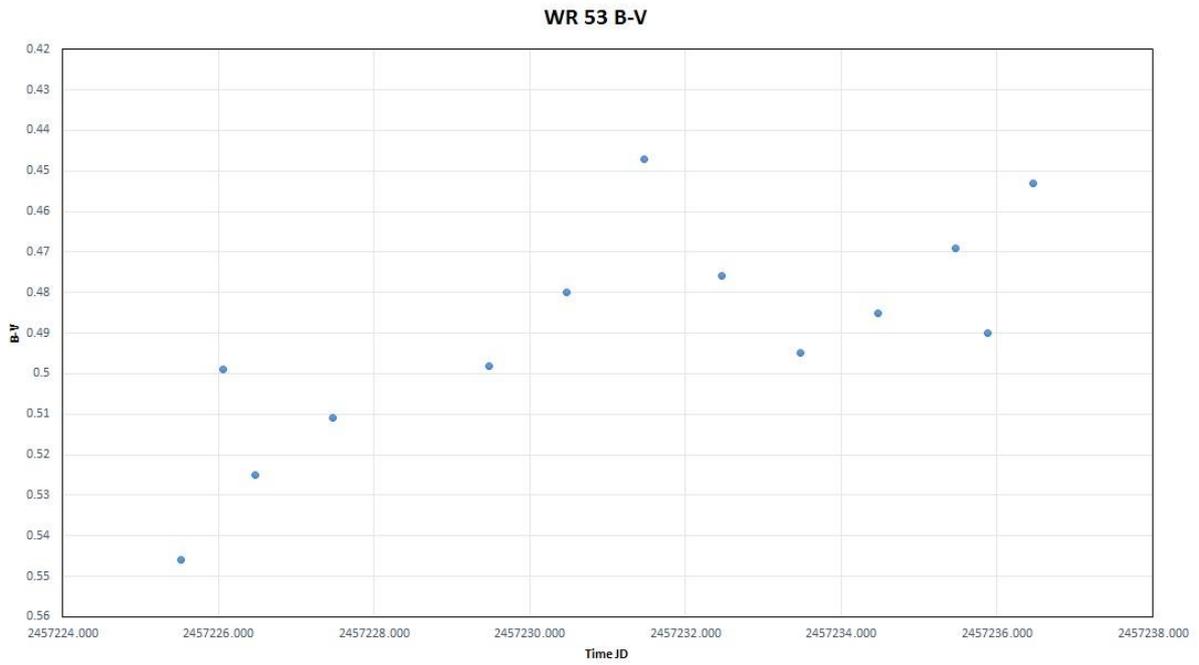

Figure 2. Light curve of WR 53 (B-V). AAVSO data from JD 2457225.4733 through JD 2457236.4693 (July 21, 2015 - August 1, 2015) showing a slight reddening at minimum and the rising trend.